\documentclass[prd,floatfix,onecolumn,amsmath,amssymb,floatfix]{revtex4}
\usepackage{graphicx,color,dcolumn,booktabs,bm}
\usepackage{subfigure}
\usepackage{longtable,lscape}
\usepackage{amssymb}
\usepackage{indentfirst}
\usepackage{epsfig}
\usepackage{feynmf}   
\usepackage{epstopdf}   
\usepackage{slashed}  
\usepackage{cases}
\usepackage[pdfpages]{xcolor}
\definecolor{maroon}{RGB}{139,25,150}
\usepackage{multirow}
\usepackage{float}
\usepackage{pgf}
\usepackage{graphicx,color,dcolumn,booktabs,bm}
\usepackage[colorlinks, citecolor=blue,anchorcolor=red,menucolor=red, linkcolor=red,filecolor=red,runcolor=red,urlcolor=blue,frenchlinks=red, urlcolor=blue]{hyperref}

\begin{document}

	\preprint{}
	
\title{\color{maroon}{Mechanical properties of the nucleon from the generalized parton distributions}}

\author{The MMGPDs Collaboration:\\
Muhammad Goharipour$^{1,2}$}
\email{muhammad.goharipour@ipm.ir}
\thanks{Corresponding author}

\author{Hadi Hashamipour$^{3}$}
\email{hadi.hashamipour@lnf.infn.it}

\author{H.~Fatehi$^{4}$}
\email{apranik.fatehi@cern.ch}

\author{Fatemeh Irani$^{4}$}
\email{f.irani@ut.ac.ir}

\author{K.~Azizi$^{4,5,2}$}
\email{kazem.azizi@ut.ac.ir}

\author{\\ And\\ S.V. Goloskokov$^{6}$}
\email{goloskkv@theor.jinr.ru}
	
\affiliation{
$^{1}$School of Physics, Institute for Research in Fundamental Sciences (IPM), P.O. Box  19395-5531, Tehran, Iran\\
$^{2}$School of Particles and Accelerators, Institute for Research in Fundamental Sciences (IPM), P.O. Box 19395-5746, Tehran, Iran\\
$^{3}$Istituto Nazionale di Fisica Nucleare, Gruppo collegato di Cosenza, I-87036 Arcavacata di Rende, Cosenza, Italy\\
$^{4}$Department of Physics, University of Tehran, North Karegar Avenue, Tehran 14395-547, Iran\\
$^{5}$Department of Physics, Dogus University, Dudullu-\"{U}mraniye, 34775 Istanbul, T\"urkiye \\
$^{6}$Bogoliubov Laboratory of Theoretical Physics, Joint Institute for Nuclear Research, Dubna 141980, Moscow region, Russia}

\date{\today}

\begin{abstract}

The proton's internal structure is characterized not only by its charge and magnetic distribution but also by its mechanical and mass properties, which are encoded in the energy-momentum tensor (EMT) of quantum chromodynamics (QCD). These properties provide insights into the spatial distributions of energy, pressure, and shear forces within the proton.
Understanding the proton's internal structure, including properties such as its mechanical and mass radii, is essential for unraveling the complex interplay between quarks and gluons that govern its stability and dynamics. In this study, we investigate the gravitational form factors (GFFs) of the proton, particularly the D-term, which encodes key information about the internal stress distribution, pressure, and shear forces within the nucleon. Using a model for skewness-dependent generalized parton distributions constructed from the double-distribution representation, we extract the quark contribution to the $ D(t) $ GFF of the EMT by analyzing available data on Compton form factors. We then employ this extracted GFF to explore the mechanical properties of the proton, including its mechanical and mass radii, as well as the internal pressure and shear force distributions. Our results provide new insights into the proton's internal structure and contribute to the broader understanding of nucleon properties.
	
\end{abstract}

\maketitle

\section{Introduction}\label{sec:one} 

Nucleons are the main building blocks of visible matter. Protons are stable composite particles made of quarks and gluons bounded via strong interactions, while free neutrons  decay dominantly via beta decay to protons.  Neutrons become stable when they are put together with protons and other neutrons inside the nuclei or medium with higher densities like neutron stars as compact astrophysical objects. Understanding  the properties of nucleons is one of the main objectives of quantum chromodynamics (QCD).  Increasing our information on the properties and internal structures of nucleons helps us better understand the perturbative and nonperturbative nature of QCD,  especially the confinement as one of the main open problems of physics. Indeed, our exact knowledge of the properties and interactions of nucleons with other particles can help us better understand the formation and evolution of the visible matter and scan different eras of our universe from the initial to the late time.  Our progress in this direction would even help us develop trustworthy ideas and scenarios  on other components of our universe: dark matter and dark energy.   In the matter side,  the parameters of nucleons calculated or extracted  with higher precision can not only  help us calculate the main parameters of the standard model and find solutions to anomalies recently seen and discussed at different hadron colliders,  but  also in developing successful scenarios beyond the standard model.

Although we have recorded a good progress on the understanding in the formation,  internal structure and  properties  of nucleons, there are many puzzles on their properties: puzzles of their spin,
geometric shape, different radii (electric, magnetic, mass,  mechanical, etc.), origin of their mass and distributions of different physical properties inside them and in terms of their constituents, the quarks and gluons.  Increasing our knowledge on these puzzles in order to solve them requires investigations of different interactions among different particles and currents with nucleons both experimentally and theoretically: currents such as electromagnetic, axial,  tensor and gravitonlike energy-momentum current.  In such interactions, the structure and dynamics of the QCD vacuum play an important role.  Each of these interactions is presented by some form factors (FFs) in the low-energy regime, using which helps us determine different physical quantities of the nucleons.  One of the important currents that plays good role in determination of physics of the nucleon is the energy-momentum tensor (EMT) current. Such a current that is conserved when fully considering the quark-gluon contents of the nucleons is obtained either from the QCD Lagrangian using Noether's first and second theorems and variations with respect to the involved fields---the first theorem is  associated with global  translations, while the second is associated with local spacetime translations (for details, see, for instance, \cite{Freese:2021jqs})---or by varying the action $ S_{grav} $ of QCD coupled to a weak classical torsionless gravitational background field with respect to the metric  of this curved background field~\cite{Polyakov:2018zvc}.  This procedure results again in  a symmetric conserved Belinfante-Rosenfeld EMT (for details see also~\cite{Belitsky:2005qn,Landau}).

The EMT current sandwiched between nucleonic states is decomposed in terms of four different FFs.  These FFs are called gravitational form factors (GFFs). These GFFs  are sources of  knowledge about  the fractions of the momenta carried by the partonic  constituents of the nucleon,  the energy density distribution inside it,  information about how the total angular momenta of quarks and gluons form the nucleon's total spin,  the distribution and stabilization of the strong force in the nucleon by the help of the so-called D-term and finally  the conservation of the EMT current and its dependence on the different partonic contributions through the so-called cosmological constant form factor.  Many physical and mechanical observables  of nucleon are constructed using these GFFs: forces inside nucleon,  pressure,  surface tension,  mechanical and mass  radii,  geometric shape and so on. The  GFFs were introduced by Kobzarev and Okun in 1962 for the first time,  but they have been  investigated by many models and approaches for nucleons and hyperons now: chiral perturbation theory,  bag model,  instanton picture,  chiral quark soliton model,  dispersion relation,  Skyrme model,  lattice QCD,  instant-front form, and light-cone QCD sum rules (LCSR).  Some GFFs are now available from analyzing of the generalized parton distributions (GPDs), as well (for some stories  about the GFFs and transition GFFs, see, for instance, Refs.~\cite{Polyakov:2018zvc,Azizi:2019ytx,Ozdem:2020ieh,Azizi:2020jog,Ozdem:2022zig,Dehghan:2023ytx}, and references therein).  Using the calculated  GFFs,  many properties of nucleons,  hyperons and some mesons are also available now.  This information will lead experimental and theoretical groups to perform further investigations.

The mechanical properties of the nucleon, such as pressure and shear forces, which are encoded in the GFFs, are fundamental to understanding the internal structure and dynamics of nucleons~\cite{Shanahan:2018nnv,Shanahan:2018pib,Pefkou:2021fni,Yang:2018nqn,Lowdon:2017idv,Teryaev:2016edw,Lorce:2015lna,Kim:2012ts,Goeke:2007fp,Polyakov:2002yz,Pagels:1966zza,Lorce:2018egm,Polyakov:2018zvc,Hackett:2023rif,Burkert:2023wzr,Cao:2024zlf,Duran:2022xag,Mamo:2021krl,Guo:2021ibg,Guo:2023pqw,Guo:2023qgu,Kharzeev:2021qkd,Chakrabarti:2020kdc,More:2021stk,Dehghan:2025ncw,Azizi:2019ytx,Avelino:2019esh,Kumericki:2019ddg,Burkert:2018bqq,Choudhary:2022den,Broniowski:2025ctl,Fujii:2025aip,Song:2025zwl,Won:2023ial} as mentioned above. These properties are accessible through the EMT, which can be probed utilizing high-energy exclusive scattering experiments~\cite{Ji:1998pc,Goeke:2001tz,Belitsky:2001ns,Diehl:2003ny,Belitsky:2005qn,Guidal:2013rya,Kumericki:2016ehc,Mezrag:2022pqk,Dutrieux:2021nlz}. The study of these mechanical properties provides insights into the stability and equilibrium conditions within nucleons, revealing how quarks and gluons interact to form these particles. This understanding is crucial for both particle and hadron physics, as it helps elucidate the behavior of nuclear matter under extreme conditions, such as those found in compact stars~\cite{Weber:2004kj} as also previously mentioned.  Additionally, the above  mechanical properties of nucleons play a significant role in the development of theoretical models and simulations, contributing to our broader comprehension of the strong force and the fundamental structure of matter.

Among the GFFs, the $ D $ form factor, which is related to the D-term in the language of GPDs, is particularly significant~\cite{Lorce:2018egm,Polyakov:2018zvc,Burkert:2018bqq,Kumericki:2019ddg,Choudhary:2022den,Burkert:2023wzr,Cao:2024zlf}. It is associated with the internal stress distribution within a hadron. It can be interpreted as a measure of the forces that bind the hadron together, particularly in terms of the pressure and shear forces exerted by quarks and gluons reflecting the internal mechanical properties of the hadron.
A negative $ D $ is conjectured to be necessary for the stability of hadrons.  As noted, the GFF $ D $ can be calculated using various theoretical approaches, including holographic QCD~\cite{Fujita:2022jus,Fujii:2024rqd,Sugimoto:2025btn}, lattice QCD~\cite{Shanahan:2018nnv,Shanahan:2018pib,Pefkou:2021fni,Hackett:2023rif}, and LCSR~\cite{Azizi:2019ytx,Dehghan:2025ncw}. It can also be determined from the QCD analysis of the experimental data related to the deeply virtual Compton scattering (DVCS)~\cite{Kumericki:2019ddg,Burkert:2018bqq} and near threshold photoproductions of $ J/\psi $ particles~\cite{Guo:2021ibg,Guo:2023pqw,Guo:2023qgu,Duran:2022xag,Kharzeev:2021qkd}.

The GFFs can also be interpreted as moments of GPDs~\cite{Muller:1994ses,Radyushkin:1996nd,Ji:1996nm,Ji:1996ek,Burkardt:2000za} in the ERBL region. The D-term is a contribution to the unpolarized GPDs which determines the asymptotic of them in the limit of renormalization scale $ \mu \rightarrow \infty $~\cite{Goeke:2001tz,Polyakov:2018zvc}. In the context of the hard-exclusive reactions, the D-term emerges as a subtraction constant in fixed-$ t $ dispersion relations~\cite{Kumericki:2019ddg,Burkert:2018bqq,Anikin:2007yh,Diehl:2007jb}.

Having the GFF $ D $  in hand, one can calculate the mechanical properties of the nucleons such as their mechanical and mass radii as well as the pressure and shear forces inside the nucleon~\cite{Polyakov:2018zvc}. In the present study, we first construct a model for the skewness-dependent GPDs based on the double-distribution (DD) representation~\cite{Radyushkin:1997ki,Radyushkin:1998bz,Radyushkin:1998es,Musatov:1999xp,Polyakov:1999gs} and using the zero-skewness GPDs obtained from the analysis of the elastic scattering processes~\cite{Hashamipour:2022noy} in Sec.~\ref{sec:two}. Then, we phenomenologically determine the GFF $ D^Q $ (or $ d_1^Q $), where $ Q $ denotes the contribution of quarks to the proton GFF $ D $, by analyzing the available data for the Compton FFs (CFFs) in Sec.~\ref{sec:three}. As the next step, we study the mechanical properties of the proton using the extracted GFF $ D $ in Sec.~\ref{sec:four}. Finally, we summarize our results and conclusions in Sec.~\ref{sec:five}.

\section{The skewness-dependent GPDs}\label{sec:two}

GPDs play a crucial role in particle physics, particularly in understanding the structure of hadrons such as protons and neutrons~\cite{Diehl:2003ny,Boffi:2007yc,Diehl:2015uka,Belitsky:2005qn,Ji:2004gf}. Representing correlations between the transverse
position and the longitudinal momentum of partons, GPDs provide a framework for three-dimensional (3D) imaging (tomography) of quark and gluon distributions within hadrons. 
This provides a detailed understanding of the spatial distribution of these particles and insights into their dynamics within hadrons. Moreover, they help in understanding how quarks and gluons contribute to the total spin of the nucleon~\cite{Cichy:2024afd}. GPDs are accessed through hard-exclusive processes such as DVCS and deeply virtual meson production (DVMP)~\cite{Ji:1998pc,Goeke:2001tz,Belitsky:2001ns,Diehl:2003ny,Belitsky:2005qn,Guidal:2013rya,Kumericki:2016ehc,Mezrag:2022pqk,Dutrieux:2021nlz} (see also Ref.~\cite{Hashamipour:2022noy} and references therein) or single diffractive hard-exclusive processes~\cite{Qiu:2022pla}. Another importance of GPDs is that 
their different Mellin moments are associated with
different hadron FFs including the electromagnetic FFs, axial FFs, GFFs, and transition FFs~\cite{Muller:1994ses,Radyushkin:1996nd,Ji:1996nm,Ji:1996ek,Burkardt:2000za,Bernard:2001rs,Pagels:1966zza,Guidal:2004nd}. This makes GPDs also an essential ingredient of different
types of elastic scattering processes.

In general, GPDs are nonperturbative objects that depend
on four variables: $ x $, $ \mu^2 $, $ \xi $, and $ t $.  The variables $ x $ and $ \mu $ are exactly those which describe the deep inelastic scattering (DIS) process~\cite{Blumlein:2012bf} and are involved in ordinary unpolarized (polarized) parton distribution functions (PDFs) $ q(x,\mu^2) $ ($ \Delta q $), i.e, the longitudinal momentum fraction of the nucleon carried by partons and the factorization scale at which the partons are resolved, respectively. In processes where the proton remains intact during the scattering (including different
types of hard-exclusive and elastic scattering processes), two extra variables are needed to describe the process completely: $ \xi $ which is the longitudinal momentum transfer, known as skewness, and $ t $ which is the negative momentum transfer squared. In the so-called forward limit, $ \xi=0 $ and $ t=0 $, GPDs are reduced to PDFs which give information on only the longitudinal hadron structure, while GPDs involve information in the transverse plane too. 

There are various approaches to model GPDs considering the fact that calculating them from first principles of QCD is quite difficult. In recent years, there have been great efforts to calculate GPDs and their moments from 
lattice QCD simulations~\cite{Hagler:2003jd,Gockeler:2003jfa,LHPC:2007blg,Constantinou:2020hdm,Alexandrou:2021bbo,Lin:2021brq,Riberdy:2023awf,Bhattacharya:2023nmv,Cichy:2023dgk,Holligan:2023jqh,Bhattacharya:2023jsc,Bhattacharya:2024qpp}. A direct calculation using effective quark models, such as the bag model~\cite{Ji:1997gm,Tezgin:2024tfh}, the chiral quark-soliton model~\cite{Diakonov:1987ty,Petrov:1998kf,Wakamatsu:2005vk,Ossmann:2004bp,Kim:2024ibz,Nematollahi:2024wrj}, the light-front Hamiltonian approach~\cite{Mukherjee:2002pq,Mukherjee:2002xi,Thakuria:2024nyv,Kaur:2023lun,Liu:2024umn}, the nonrelativistic and light-cone constituent quark model~\cite{Scopetta:2003et,Scopetta:2004wt,Boffi:2002yy,Pasquini:2005dk,Luan:2023lmt,Luan:2024vgv,Luan:2025nnm}, the nonlocal chiral quark model~\cite{Son:2024uet,Gao:2024ajz,He:2022leb}, the meson-cloud model~\cite{Pasquini:2006dv,Pasquini:2006ib}, the aligned-jet model~\cite{Freund:2002qf,Khanpour:2017slc}, the large-momentum effective theory~\cite{Ji:2014gla,Ji:2020ect}, the light-front quark-diquark (LFQD) model~\cite{Mondal:2015uha}, and the light-front spectator model~\cite{Sain:2025kup,Thakuria:2025lcz}, is another approach to calculate GPDs or their moments. Utilizing a phenomenological approach, GPDs can also be obtained by analyzing the available experimental  or lattice data in both zero~\cite{Hashamipour:2022noy,Diehl:2004cx,Diehl:2013xca,Gonzalez-Hernandez:2012xap,Selyugin:2014sca,Ahmady:2021qed,Arami:2024qsu,Hashamipour:2019pgy,Hashamipour:2020kip,Hashamipour:2021kes,Irani:2023lol,Goharipour:2024atx,Goharipour:2024mbk,Guo:2022upw,Singireddy:2025biy} and nonzero skewness~\cite{Kumericki:2016ehc,Goloskokov:2005sd,Goloskokov:2007nt,Goloskokov:2009ia,Goloskokov:2011rd,Kroll:2012sm,Goloskokov:2022mdn,Goloskokov:2024egn,Kumericki:2006xx,Kumericki:2007sa,Kumericki:2009uq,Lautenschlager:2013uya,Muller:2013jur,Berthou:2015oaw,Moutarde:2018kwr,Kriesten:2019jep,Cuic:2020iwt,Kriesten:2021sqc,Guo:2023ahv,Guo:2024wxy,Mamo:2024jwp,Cuic:2023mki,Mamo:2024vjh} cases.

Among the phenomenological approaches, the DD representation which is based on parametrizing the hadronic matrix elements (which define GPDs) in terms of double distributions~\cite{Radyushkin:1997ki,Radyushkin:1998bz,Radyushkin:1998es,Musatov:1999xp,Polyakov:1999gs} has been successfully used for phenomenological analyses of the DVCS and DVMP data~\cite{Goloskokov:2005sd,Goloskokov:2007nt,Goloskokov:2009ia,Goloskokov:2011rd,Kroll:2012sm,Goloskokov:2022mdn,Goloskokov:2024egn}. By assuming a factorized form with a $ t $-dependent part constrained from data and a $ t $-independent part given in terms of double distributions, the DD representation has an advantage to relate GPDs to other types of parton distributions, such as PDFs and transverse momentum-dependent distributions (TMDs). The DD representation also provides a systematic way to construct GPDs that satisfy the polynomiality condition and other theoretical constraints~\cite{Polyakov:2009xir,Mezrag:2023nkp,Radyushkin:2023ref}. The polynomiality property of GPDs states that the Mellin moments of GPDs are polynomials in the skewness parameter $ \xi $. 

According to the DD representation, the skewness-dependent GPD $ H(x,\xi,t) $ (both quark and gluon distributions), for example, can be expressed as an integral over $ t $-dependent double distribution $ f(\beta, \alpha, t) $
\begin{equation}
\label{Eq1}
 H_{DD}(x, \xi, t) = \int_{-1}^{1} d\beta \int_{-1+|\beta|}^{1-|\beta|} d\alpha\, \delta(x - \beta - \xi \alpha)\, f(\beta, \alpha, t)\,,
\end{equation}
where $ f(\beta, \alpha, t) $ for each parton $ i $ can be considered as the product of a $ t $-dependent part with a Regge behavior~\cite{Goloskokov:2005sd,Goloskokov:2007nt} and a $ t $-independent DD which can be written as
\begin{equation}
\label{Eq2}
F_i(\beta, \alpha)= h(\beta, \alpha)\,q_i(\beta)\,,
\end{equation}
where $ q_i(\beta) $ is the forward PDF and the proﬁle function $ h_i(\beta, \alpha) $ is parametrized as~\cite{Musatov:1999xp}
\begin{equation}
\label{Eq3}
h_i(\beta, \alpha)=\frac{\Gamma(2n_i+2)}{2^{2n_i+1}\,\Gamma^2(n_i+1)}
                   \,\frac{[(1-|\beta|)^2-\alpha^2]^{n_i}}
                           {(1-|\beta|)^{2n_i+1}}\,,\end{equation}
where parameter $ b $ characterizes the strength of the $ \xi $ dependence of the GPDs. Although it can be considered as a free parameter in the extraction of GPDs from the data of hard-exclusive processes, it has been shown that $  n_{\mathrm{val}}=1 $ for valence quarks and $  n_{\mathrm{sea}}=n_g =2$ for the sea quarks and gluon are good choices to describe the data~\cite{Goloskokov:2007nt}. Note that in the forward limit, $ \xi,t\rightarrow 0 $, GPDs reduce to ordinary PDFs $ q_i(x) $.

The $ x $ and $ t $ dependence of GPDs at zero-skewness limit can be well determined by analyzing experimental data from different types of elastic scattering processes including the elastic electron-nucleon scattering, elastic (anti)neutrino-nucleon scattering, and the wide-angle Compton scattering (WACS)~\cite{Hashamipour:2022noy,Hashamipour:2019pgy,Hashamipour:2020kip,Hashamipour:2021kes,Irani:2023lol,Goharipour:2024atx,Goharipour:2024mbk}. The idea is to use an ansatz for zero-skewness GPDs as follows~\cite{Diehl:2004cx,Diehl:2013xca} 
\begin{equation}
{\cal F}_i (x,t)= q_i(x)\exp [tf_i(x)]\,,
\label{Eq4}
\end{equation}
which relates GPDs $ {\cal F}_i $ to forward PDFs $ q_i(x) $ at a certain renormalization scale $ \mu $. For example, in Ref.~\cite{Hashamipour:2022noy}, the unpolarized and polarized PDFs have been taken from the next-to-leading-order (NLO) \texttt{NNPDF} analyses~\cite{NNPDF:2021njg,Nocera:2014gqa} at $ \mu=2 $ GeV utilizing the \texttt{LHAPDF} package~\cite{Buckley:2014ana} . The profile function $ f_i(x) $ in Eq.~(\ref{Eq4}), which controls the $ t $ dependence of GPDs at different regions of $ x $, can be parametrized as 
\begin{equation}
\label{Eq5}
f_i(x)=\alpha^{\prime}(1-x)^3\log\frac{1}{x}+B(1-x)^3 + Ax(1-x)^2\,,
\end{equation}
where the unknown parameters $ \alpha^{\prime} $, $ A $, and $ B $ for each quark flavor (note that such analyses do not usually contain the gluon contribution) are obtained from the $ \chi^2 $ analysis of the related experimental data. Although such an ansatz is very successful to describe the elastic scattering data, the resulted GPDs cannot be used to describe hard-exclusive processes such as DVCS and DVMP. A novel idea in this context is to combine the zero-skewness GPDs $ {\cal F}_i (x,t) $ obtained, i.e., from the analysis of Ref.~\cite{Hashamipour:2022noy} and the $ t $-independent DD $ F_i(\beta, \alpha) $ of Eq.~(\ref{Eq2}) to construct $ t $-dependent DD $ f_i(\beta, \alpha, t) $ in Eq.(\ref{Eq1}). This leads to the $ \xi $-dependent GPDs $ H_{DD}(x, \xi, t) $ with hard constraints on the zero-skewness limit (in particular on the $ t $ dependence of GPDs). Note that the measurements of the hard-exclusive processes like DVCS and DVMP are corresponding to the small $ t $ values. So, the extracted $ \xi $-dependent GPDs from the analysis of DVCS and DVMP data are just well constrained at small $ t $ regions.

Although Eq.~(\ref{Eq1}) provides the $ \xi $ dependence of GPDs well, it does not satisfy the polynomiality property of GPDs~\cite{Ji:1998pc,Polyakov:1999gs}. Including a D-term as follows, the DD representation satisfies polynomiality~\cite{Diehl:2003ny,Boffi:2007yc,Belitsky:2005qn,Ji:2004gf}. For the quark distributions, the final GPDs will be as
\begin{equation}
\label{Eq6}
H^q(x, \xi, t)= H_{DD}^q(x, \xi, t) + \theta(\left| \xi \right|-\left| x \right|)D^q(\frac{x}{\xi},t)\,.
\end{equation}

Reviewing some important points about D-term would be worthwhile in this context~\cite{Diehl:2003ny,Boffi:2007yc}. First of all, note that the D-term contributions to GPDs have support only in the ERBL region $ \left| x \right| \leq  \left| \xi \right|$, so that it is invisible in the forward limit. Another point should be mentioned is that the D-term does not contribute to the nonsinglet distributions and contributes only to the singlet-quark and gluon distributions. Moreover, it contributes to either unpolarized GPDs $ H $ and $ E $, but there is no analog of the D-term for the helicity dependent GPDs $ \widetilde{H} $ and $ \widetilde{E} $. Note also that the same function $ D^{q,g} $ contributes with opposite sign to $ H^{q,g} $ and to $ E^{q,g} $. Finally, the D-term does not change remarkably under evolution, so that its effect under evolution is at the level of a few percent.

Considering $ z=x/\xi $, $ D(z,t) $ have the following expansion in the Gegenbauer polynomials $ C_n^{3/2}(z) $~\cite{Diehl:2003ny,Burkert:2023wzr,Polyakov:2018zvc,Goeke:2001tz} 
\begin{equation}
\label{Eq7}
D^q(z,t)= (1 - z^2) \sum_{\text{odd}~n}^\infty d_n^q(t)\, C_n^{3/2}(z) \,.
\end{equation}
In the limit of $ \mu\rightarrow \infty $ (note that the renormalization scale dependence has not been indicated in the above formulas), all $ d_n^q(t) $ go to zero except $ d_1^q(t) $,  which is related to the GFF $ D^q(t) $ of the EMT as follows
\begin{equation}
\label{Eq8}
D^q(t)= \frac{4}{5}\, d_1^q(t)= \int_{-1}^1 dz\,z\, D^q(z,t) \,.
\end{equation}
Now, if we consider that the flavor singlet combination of the quark D-term is dominant~\cite{Polyakov:2018zvc}, $ d_1^u\approx d_1^d\approx d_1^Q/2 $, we can determine the EMT form factor $ D^Q= D^u(t) + D^d(t)$ (neglecting effects of strange and heavier quarks) by analyzing data related to the $ \xi $-dependent GPDs of of Eq.~(\ref{Eq6}) like the CFF data (see the following).

\begin{figure}[!htb]
    \centering
\includegraphics[scale=0.5]{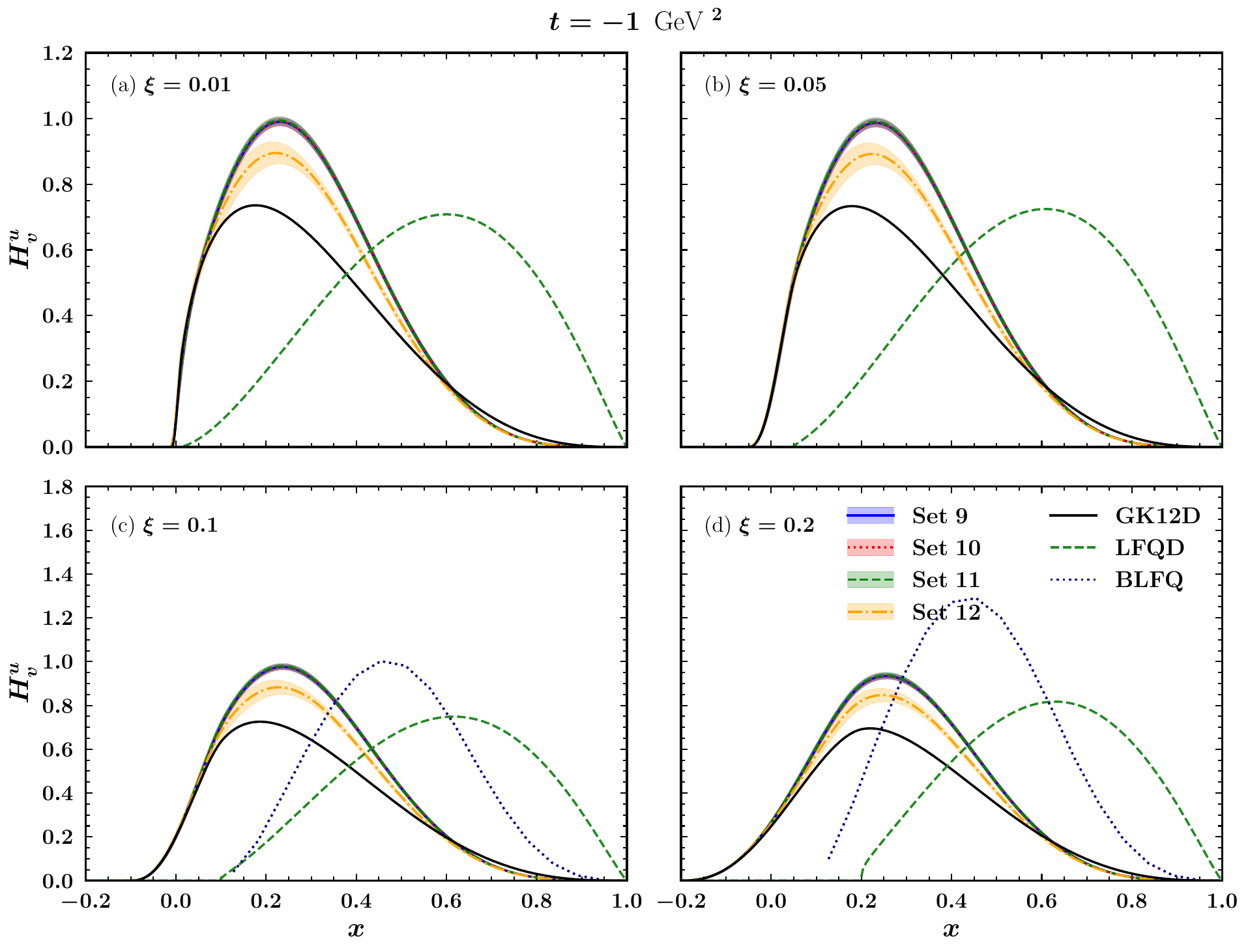}    
    \caption{Comparison of the unpolarized up valence GPD \( H_{DD}^{u_v}(x,\xi,t) \) at \( t = -1 \) GeV\(^2\) for four values of \( \xi \) (\( \xi = 0.01, 0.05, 0.1, 0.2 \)). The results are calculated using Eq.~(\ref{Eq1}) with zero-skewness GPD sets from Ref.~\cite{Hashamipour:2022noy} (HGAG23), alongside the GK12D analysis~\cite{Kroll:2012sm}, BLFQ~\cite{Liu:2024umn} and LFQD~\cite{Mondal:2015uha} calculations. See the text for more details.}
\label{fig:Huv}
\end{figure}

Figure~\ref{fig:Huv} presents a comparison of results for the unpolarized up valence GPD \( H_{DD}^{u_v}(x,\xi,t) \), including its uncertainty calculated using Eq.~(\ref{Eq1}), with various zero-skewness GPD sets from Ref.~\cite{Hashamipour:2022noy} (referred to as HGAG23). The comparison is shown at \( t = -1 \) GeV\(^2\) and four values of \( \xi \): \( \xi = 0.01, 0.05, 0.1, 0.2 \). Additionally, the result from the GK12D analysis~\cite{Kroll:2012sm}, derived from hard-exclusive data, is included utilizing the Gepard package~\cite{Kumericki:2006xx,Kumericki:2007sa,Kumericki:2009uq,Cuic:2023mki}. Note that GK12D GPDs incorporate the D-term (although it does not matter for the valence distributions which are nonsinglet and, hence, do not include D-term). 
For further comparison, we have also included the results obtained from two theoretical approaches, namely the basis light-front quantization (BLFQ) framework~\cite{Liu:2024umn} and the LFQD model~\cite{Mondal:2015uha}. Among the GPD sets from Ref.~\cite{Hashamipour:2022noy}, Set 12 exhibits distinct behavior, while the other sets show strong agreement. This discrepancy in Set 12 arises from the exclusion of the proton ($ p $) magnetic form factor \( G_M^p \) data due to observed tensions with WACS data. However, Ref.~\cite{Goharipour:2024atx} highlights that Set 12 provides a poorer description of the JLab GMp12 data~\cite{Christy:2021snt} for the electron-proton scattering reduced cross section. The figure also reveals a notable difference between the GPDs from the GK12D analysis and those constructed using Eq.~(\ref{Eq1}) with the zero-skewness GPDs of the HGAG23 analysis. In addition, note that the results obtained from theoretical approaches tend more to the large-$ x $ region compared with the phenomenological analyses. Therefore, it is of great interest to explore the applicability and utility of the new \( \xi \)-dependent GPDs introduced here
in hadron and particle physics and related subjects.

\begin{figure}[!htb]
    \centering
\includegraphics[scale=0.5]{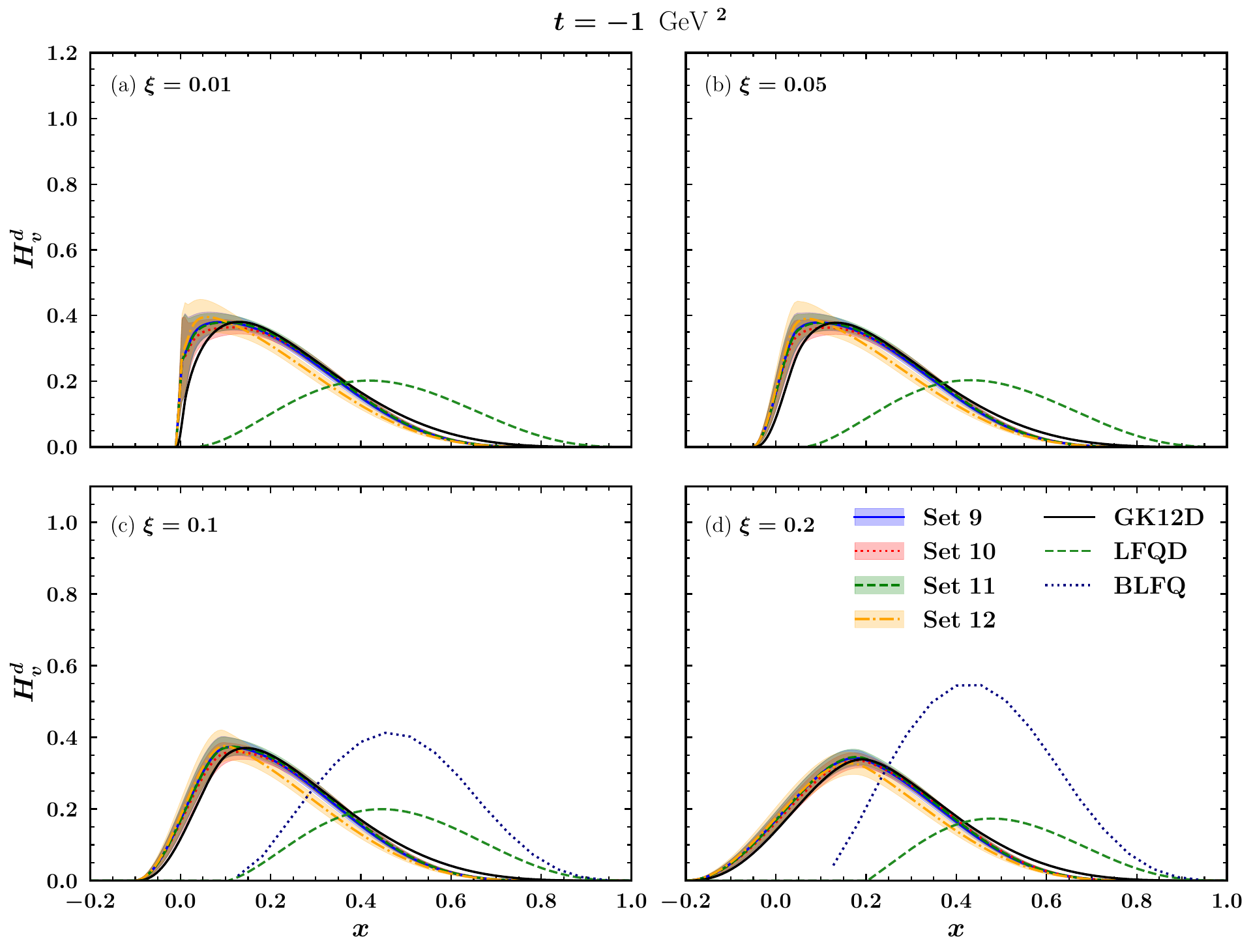}    
    \caption{The same as Fig.~\ref{fig:Huv} but for the unpolarized down valence GPD $ H_{DD}^{d_v}(x,\xi,t) $.}
\label{fig:Hdv}
\end{figure}

Figure.~\ref{fig:Hdv} shows the same results as Fig.~\ref{fig:Huv} but for the unpolarized down valence GPD $ H_{DD}^{d_v}(x,\xi,t) $. In this case, all results are in good consistency with each other. Considering Figs.~\ref{fig:Huv} and ~\ref{fig:Hdv}, one can conclude that the analysis of the elastic scattering data leads to the same down distribution but larger up distribution compared with the analysis of the data of hard-exclusive processes. So, it would be interesting to see how the differences observed affect various phenomenological studies in this field. One of them could be the determination of D-term (or the GFF $ D $) itself. Note also that the theoretical results obtained from BLFQ and LFQD are again in significant inconsistency with the phenomenological results. This highlights the need for further theoretical, experimental, and phenomenological efforts to shed light on this issue and resolve the discrepancy  observed. 

According to the above explanations, in order to have complete GPDs of Eq.~(\ref{Eq6}), we need to have $ d_1^q(t) $  as a function of $ t $. It is well established now that $ d_1^Q(t) $ (or $ D^Q $) can be determined through the analysis of the data of the DVCS observables utilizing a simple and suitable functional form and considering the fixed-$ t $ dispersion relation~\cite{Kumericki:2019ddg,Burkert:2018bqq,Dutrieux:2021nlz}. The dispersion relation relates the real and imaginary parts of CFFs through a real subtraction term~\cite{Anikin:2007yh,Diehl:2007jb}. Now, having the DD part of the  $ \xi $-dependent GPDs, i.e. $ H_{DD}(x,\xi,t) $ in Eq.~(\ref{Eq1}), which is calculable utilizing the zero-skewness GPDs of Ref.~\cite{Hashamipour:2022noy} (see the above explanations), we can determine $ D^Q(t) $ (or D-term) through the analysis of the CFFs data considering the fact that the real and imaginary parts of CFFs are related to GPDs. This is exactly what we are going to do in the present study.

\section{Determination of D-term}\label{sec:three}

As mentioned before, the GFF $ D(t) $ is a fundamental quantity that provides valuable information about the internal structure and mechanical properties of protons and other hadrons. It is essential for understanding the distribution of pressure and shear forces, contributes to the three-dimensional imaging of hadrons through its relation to GPDs, and plays a significant role in both experimental and theoretical studies of QCD~\cite{Lorce:2018egm,Polyakov:2018zvc,Burkert:2018bqq,Kumericki:2019ddg,Choudhary:2022den,Burkert:2023wzr,Cao:2024zlf}.

In this section, we are going to determine the GFF $ D^Q(t)=4/5\, d_1^Q(t) $  by performing a $ \chi^2 $ analysis of the available data for the real part of the CFFs. In this regard, we first introduce our phenomenological framework. Then, we present the result obtained for $ D^Q(t) $ and compare it with the corresponding ones obtained from other studies including the lattice and LCSR calculations.

\subsection{Phenomenological framework}

The framework of GPDs provides a way to gather information about the mechanical properties of nucleons. However, analyzing GPDs through DVCS data presents challenges. This is because the GPDs are formulated in terms of integrals over the variable $ x $, which range from $ -1 $ to 1 and are related to an underlying quark loop. Since $ x $ cannot be measured directly, expressing the DVCS cross section in terms of CFFs, which can be measured in experiments, would be helpful. Thanks to the factorization theorem, we can describe CFFs as combinations of GPDs and coefficient functions, which can be calculated at any order in perturbative QCD (pQCD).   

At leading order and leading twist, the relationship for the CFF $ {\cal H} $ related to the GPD $ H $, for example, is expressed as  
\begin{equation}
\label{Eq9}  
{\cal H}(\xi, t) = \sum_q e_q^2 \int_{-1}^{1} d x \, C^{-} H^q(x, \xi, t)\,.
\end{equation}
In this expression, the summation runs over all quark flavors. The coefficient function at leading twist is scale independent and is given by~\cite{Belitsky:2001ns} 
\begin{equation}
\label{Eq10}  
C^{\mp}(x, \xi) = \left( \frac{1}{\xi - x - i0} \mp \frac{1}{\xi + x - i0} \right)\,,  
\end{equation}
where $ e_q $ denotes the quark charge in units of the proton charge. Using the residue theorem, the integral in Eq.~(\ref{Eq9}) can be decomposed so that each CFF consists of two parts. Specifically, for the CFF $ {\cal H} $, it can be written as 
\begin{equation} 
\label{Eq11}   
{\cal H} (\xi, t) = \mathrm{Re}{\cal H}(\xi, t) + i \mathrm{Im}{\cal H}(\xi, t)\,,
\end{equation}
with the real and imaginary components defined, respectively, as follows: 
\begin{equation} 
\label{Eq12}   
\mathrm{Re}{\cal H}(\xi, t) = \sum_q e_q^2\, {\cal P} \int_{0}^{1} d x \, C^- \left[ H^q(x, \xi, t) - H^q(-x, \xi, t) \right]\,,  
\end{equation}
\begin{equation}  
\label{Eq13}   
\mathrm{Im}{\cal H}(\xi, t) = \pi \sum_q e_q^2 \left[ H^q(\xi, \xi, t) - H^q(-\xi, \xi, t) \right]\,.  
\end{equation} 

In Eq.~(\ref{Eq12}), $ {\cal P} $ indicates the Cauchy principal value, and the skewness $ \xi $ is related to the Bjorken variable $ x_B $ at leading twist by the relation $ \xi \approx \frac{x_B}{2-x_B} $.  
Furthermore, the coefficient function can be presented as  
\begin{equation}  
\label{Eq14}   
C^{\mp}(x, \xi) = \left( \frac{1}{\xi - x} \mp \frac{1}{\xi + x} \right)\,.  
\end{equation} 

There are similar expressions for the CFFs $ {\cal E} $, $ {\cal \tilde{H}} $, and $ {\cal \tilde{E}} $ correspond to the GPDs $ E $, $ \tilde{H}$, and $ \tilde{E} $, respectively. The upper sign in Eq.~(\ref{Eq14}) is employed for the unpolarized GPDs ($ H $, $ E $), while the lower sign applies to the polarized GPDs ($ \tilde{H} $, $ \tilde{E} $)~\cite{CaleroDiaz:2023orz,Moutarde:2009fg,Guidal:2008ie}.  

Considering Eqs.~(\ref{Eq6}), (\ref{Eq7}), and (\ref{Eq12}), one can determine the GFF $ d_1^Q(t) $ (or $ D^Q $) and, hence, the D-term in GPDs by analyzing the available data of the real part of CFF  $ {\cal H} $. Note that the D-term does not contribute to the imaginary part of $ {\cal H} $ for which we have $ x=\xi $, because it becomes zero at $ z=1 $ in Eq.~(\ref{Eq7}). Now, let us rewrite Eq.~(\ref{Eq12}) by putting the complete GPD of Eq.~(\ref{Eq6}) into it.
We will have 
\begin{equation}  
\label{Eq15}   
\mathrm{Re}{\cal H}(\xi, t) = \sum_q e_q^2\, {\cal P} \int_{0}^{1} d x \, C^- \left[ H_{DD}^q(x, \xi, t) + H_{DD}^{\bar{q}}(x, \xi, t) + \theta(\left| \xi \right|-\left| x \right|) D^q \left(\frac{x}{\xi},t \right) \right ]\,,  
\end{equation} 
where one of the properties of GPDs, i.e. $ H^q(-x,\xi, t)=-H^{\bar{q}}(x,\xi, t) $, has been used to include the sea quark contributions. Note also, as mentioned before, the D-term contributes in only the singlet distributions (like the sum of quark and antiquark distributions here). According to the explanations presented under Eq.~(\ref{Eq8}), considering $ d_1^u\approx d_1^d\approx d_1^Q/2 $, the above equation is expressed as
\begin{eqnarray}
\label{Eq16}  
\nonumber 
\mathrm{Re}{\cal H}(\xi, t) &=& \sum_q e_q^2\, {\cal P} \int_{0}^{1} d x \, C^- \left[ H_{DD}^q(x, \xi, t) + H_{DD}^{\bar{q}}(x, \xi, t) \right ]\\
&+& \frac{5}{18}\, d_1^Q(t)\, {\cal P} \int_{0}^{1} d x \, C^- \theta \left(\left| \xi \right|-\left| x \right|\right) \left(1-(\frac{x}{\xi})^2 \right) C_1^{3/2}\left(x/\xi \right)\,, 
\end{eqnarray}  
where  $ C_1^{3/2}\left(z \right)= 3z$. Now, if we calculate $ H_{DD}^q(x, \xi, t) + H_{DD}^{\bar{q}}(x, \xi, t)  $ using Eq.~(\ref{Eq1}) with the zero-skewness GPDs of Ref.~\cite{Hashamipour:2022noy} (we take Set 11 that has been obtained by analyzing more experimental data), we can determine $ d_1^Q(t) $ (and, hence, $ D^Q $) from the $ \chi^2 $ analysis of the data of the real part of the CFFs.

To achieve this goal, we utilize the data presented in Refs.~\cite{Burkert:2018bqq,Moutarde:2009fg}, which have been extracted from the original data of DVCS observables. The dataset from Ref.~\cite{Burkert:2018bqq}, referred to as BGE18, covers the ranges \( -0.15 < t < -0.34 \) and \( 0.067 < \xi < 0.223 \). The dataset from Ref.~\cite{Moutarde:2009fg}, referred to as JLab09, spans a wider range in \( t \), specifically \( -0.13 < t < -0.76 \), with a similar coverage in \( \xi \), \( 0.071 < \xi < 0.22 \).

For the functional form of the $ d_1^Q(t) $ we use the same form used in the analysis of Refs.\cite{Burkert:2018bqq,Kumericki:2019ddg}, i.e,
\begin{equation}  
\label{Eq17}   
 d_1^Q(t)= d_1^Q(0)\left(1 - \frac{t}{M^2} \right)^{-\alpha}\,,
\end{equation} 
where $ d_1^Q(0) $, $ M^2 $, and $ \alpha $ are the parameters that can be determined from the fit to data.

\subsection{Results}

In this subsection, we are going to determine the EMT form factor $ D^Q(t)=4/5\,d_1^Q(t) $ by performing a $ \chi^2 $ analysis of the CFFs data and utilizing the phenomenological framework introduced in the previous subsection. To this aim, we consider $ \alpha=3 $ in Eq.~(\ref{Eq17}) as suggested in Ref.~\cite{Burkert:2018bqq} which is consistent with the asymptotic behavior
required by the dimensional counting rules in QCD~\cite{Lepage:1980fj}. Moreover, since there are not enough constraints from CFF data on the $ t $ behavior of $ d_1^Q(t) $, we fix the parameter $ M^2 $ during the fit. Actually, taking $ M^2 $ as a free parameter, leads to a large value for it which, in turn, makes $ d_1^Q(t) $ unphysical. In order to investigate the impact of $ M^2 $ on the final results, we scan its value by changing it in the interval $ 1 < M^2 < 3 $.

\begin{figure}[!htb]
    \centering
\includegraphics[scale=0.8]{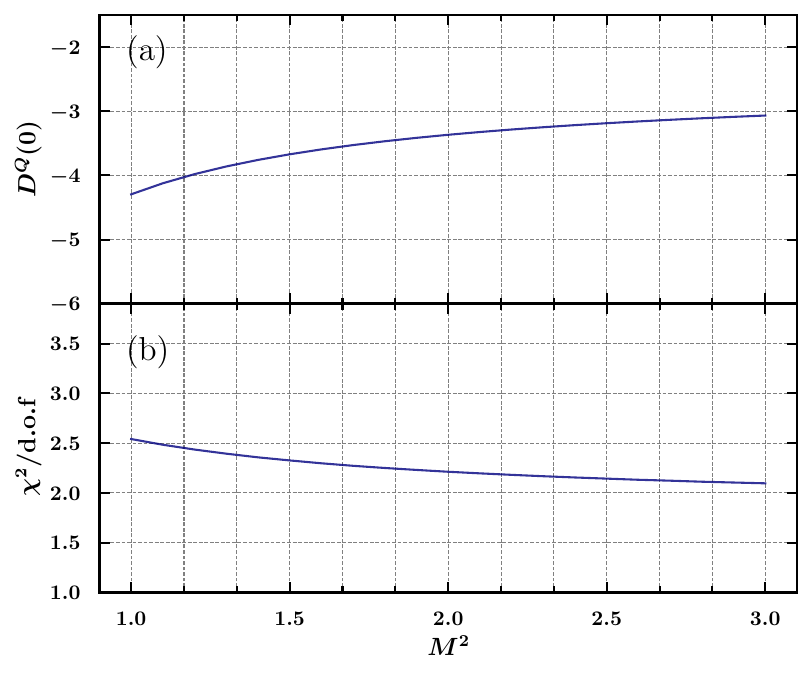}    
    \caption{(a) $ D^Q(0) $ in terms of $ M^2 $. (b) $\chi^2 /\mathrm{d.o.f.} $ in terms of $ M^2 $. See the text for more details.}
\label{fig:M2Scan}
\end{figure}

Figure~\ref{fig:M2Scan} (top panel) shows the results of $ D^Q(0) $ in terms of $ M^2 $ values. As can be seen, as the value of $ M^2 $ increases, the value obtained for $ D^Q(0) $ from the fit to data becomes smaller in magnitude so that for $ M^2=1 $ GeV$ ^2 $ and $ M^2=3 $ GeV$ ^2 $ one obtains $ D^Q(0)=-4.29 $ and $ D^Q(0)=-3.07 $, respectively. It is also of interest to see how the values of $ \chi^2 $ change when we fix $ M^2 $ on different values between 1 and 3. The bottom panel in Fig.~\ref{fig:M2Scan} shows the results of the values of total $ \chi^2 $ divided by the number of degrees of freedom, $\chi^2 /\mathrm{d.o.f.} $, in terms of $ M^2 $ values. This figure clearly indicates that as the value of $ M^2 $ increases, the changes in the value of $ \chi^2 $ becomes more insignificant. To be more precise, after $ M^2=2.3 $ GeV$ ^2 $ we have $ \Delta \chi^2 < 1 $. 

\begin{figure}[!htb]
    \centering
\includegraphics[scale=1.0]{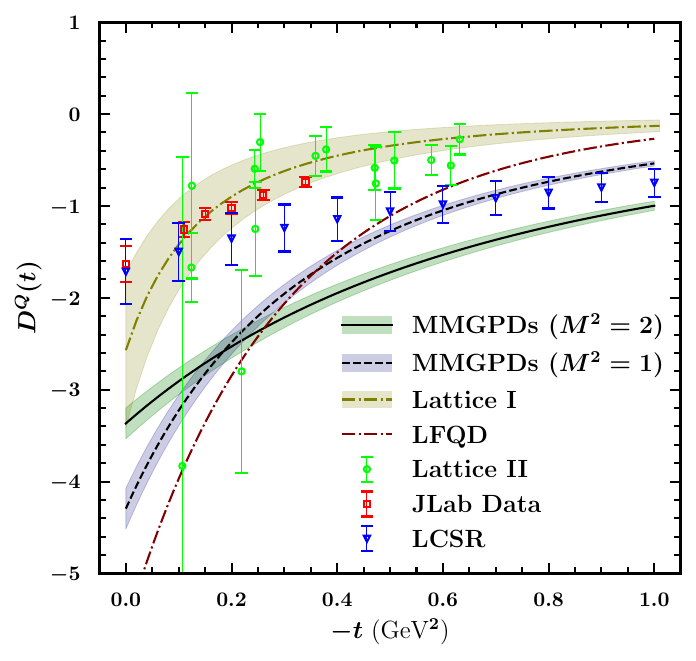}    
    \caption{A comparison between the results obtained for $ D^Q(t)=4/5\, d_1^Q $ from the present study (MMGPDs) at $ \mu^2=4 $ GeV$ ^2 $ and the corresponding ones from the lattice QCD~\cite{Hackett:2023rif,LHPC:2007blg}, LCSR~\cite{Azizi:2019ytx}, and light-front quark-diquark model calculations~\cite{Chakrabarti:2020kdc} as well as the phenomenological analysis of JLab data~\cite{Burkert:2018bqq} at $ \mu^2=1.5 $ GeV$ ^2 $.}
\label{fig:DQt}
\end{figure}

Considering $ M^2=2 $ GeV$ ^2 $, the values of $ \chi^2 $ per number of data points, $\chi^2$/$ N_{\textrm{pts.}} $, for BGE18 and JLab09 datasets are $ 127.71/50 $ and $ 64.77/38 $, respectively, which leads to $\chi^2 /\mathrm{d.o.f.}=2.21 $. Moreover, for the optimum value of $ D^Q(0) $ and its uncertainty we have $ D^Q(0)=-3.37 \pm 0.17 $. Figure~\ref{fig:DQt} shows a comparison between the results obtained for $ D^Q(t)=4/5\, d_1^Q $ from the present study at $ \mu^2=4 $ GeV$ ^2 $ and two values of $ M^2 $, namely $ M^2=1,2 $ GeV$ ^2 $, and the corresponding ones from recent~\cite{Hackett:2023rif} and old~\cite{LHPC:2007blg} lattice QCD calculations (labeled as Lattice I and Lattice II, respectively)  as well as the LCSR~\cite{Azizi:2019ytx} and light-front quark-diquark model~\cite{Chakrabarti:2020kdc} (labeled as LFQD) predictions. Moreover, we have shown the corresponding phenomenological results obtained from the analysis of Ref.~\cite{Burkert:2018bqq} at $ \mu^2=1.5 $ GeV$ ^2 $, though it has been established that by taking into account some modification, $ D^Q(t) $ becomes remarkably small~\cite{Kumericki:2019ddg}. As can be seen, our approach leads to greater values for $ D^Q(t) $ almost at all values of $ t $ compared with other studies. Note that if we choose $ M^2=1 $ GeV$ ^2 $, the magnitude of $ D^Q(0) $ even becomes larger (4.29), but $ D^Q(t) $ goes to zero faster than before as the value of $ t $ increases. Overall, our results, particularly for $ M^2=1 $ GeV$ ^2 $, show better consistency with the LFQD model.

This point should be noted again that these results obtained considering $ d_1^u= d_1^d= d_1^Q/2 $ and $ d_1^Q=d_1^u+d_1^d $ which are consistent with the definitions and assumptions made in Refs.~\cite{Polyakov:2018zvc,Burkert:2018bqq,Kumericki:2019ddg}. However, the assumption $ d_1^u= d_1^d= d_1^Q/2 $ may not be a good choice for the DVMP processes, e.g, the $ \rho $ electroproduction on the proton~\cite{Guidal:2007cw}. Actually, we know that considering D-term can increase the DVMP cross section at low values of $ W $, where $ W $ is the center of mass energy of the photon-proton system. On the other hand, the assumption $ d_1^u= d_1^d $ leads to a zero contribution, e.g, for the $ \rho^+ $ production because we have $  d_1^u-d_1^d=0 $. Note also that if we define the GFF $ d_1^Q(t) $ (the quark contributions of the EMT GFF $ d_1 $) as $ d_1^Q(t)=\sum_q e_q^2 d_1^q(t) $, it is not needed to make any assumptions of the quark decompositions and $ d_1^Q(t) $ can be directly obtained from fit to the data. If we do so, we get a smaller D-term so that the GFF $ D^Q(t) $ in Fig.~\ref{fig:DQt} becomes more consistent with the results of lattice QCD and LCSR.

According to the results obtained in this section, one can conclude that utilizing  the $ \xi $-dependent GPDs constructed with the zero-skewness GPDs obtained from the analysis of the elastic scattering data~\cite{Hashamipour:2022noy} (see Sec.~\ref{sec:two}) to determine D-term through the analysis of the CFFs data leads to a remarkable value for $ D^Q(0) $ which is in contrast with the result obtained from the analysis of the data of the DVCS observables utilizing the CFF parametrizations~\cite{Kumericki:2019ddg}. Note again that the elastic scattering processes cover a wider range of $ t $, including the larger $ t $ values, while the DVCS observables are corresponding only to the small values of $ t $. Let us clarify this issue further. As can be seen from Fig. 1, the $ \xi $-dependent GPDs constructed from elastic scattering data have a greater magnitude compared to those obtained from DVCS data.  On the other hand, 
during the fitting process, an increase in the first term of $ \mathrm{Re}{\cal H}(\xi, t)  $ (which depends on the GPDs) leads to a decrease in the second term, which is related to the GFF $ D^Q(t) $ (see Eq.~(\ref{Eq16})).  This explains why the GPDs constructed in this work result in a more negative $ D^Q(t) $ in Fig. 4 compared to the JLab data.

Considering the possible correlation between the large $ t $ and $ \xi $ regions, the results obtained here more likely predicts that the forthcoming hard-exclusive data from present (Jefferson Laboratory, COMPASS at CERN) and future (Electron-Ion Collider) facilities will be led to a more considerable EMT form factor $ D^Q $ than the available data now. 
It is also worth mentioning in this context that, although the result obtained in this section can provide good insight in the determination of the D-term (or the EMT form factor $ D^Q(t) $) using the $ \xi $-dependent GPDs constructed in Sec.~\ref{sec:two} and the approach explained in this section, the procedure should be also tested with the original DVCS data which leads, in turn, to the inclusion of more data points in the analysis and, hence, more constraints on the fit parameters. We will do this mission in our next study.

%

\section{The mechanical properties of the proton}\label{sec:four} 

As mentioned before, the mechanical properties of hadrons are in direct relation with the EMT GFF $ D(t) $. In the previous section, we determined $ D^Q(t) $ GFF of the proton (only the quark contributions) by performing a $ \chi^2 $ analysis of the data of the real part of CFFs considering its relation with the $ \xi $-dependent GPDs and utilizing the zero-skewness GPDs obtained from the analysis of Ref.~\cite{Hashamipour:2022noy}. Now, we can calculate the mechanical properties of the proton such as the pressure and shear forces as well as the mechanical and mass radii using the extracted $ D^Q(t) $.

\subsection{Pressure and shear force}

As discussed in the Introduction, understanding the internal mechanical structure of nucleons, such as pressure and shear forces, is critical to elucidate the dynamics of QCD at low energies. These quantities are encoded in the EMT and are accessible through GPDs in hard-exclusive processes as explained in Sec.~\ref{sec:two}.
From a physical point of view, the pressure $p(r)$ and shear force $s(r)$ describe the internal mechanical equilibrium of a nucleon~\cite{Shanahan:2018nnv,Lorce:2025oot,MatiasAstorga:2024ypg}. To be more precise, the pressure $p(r)$ represents the radial force per unit area at a distance $r$ from the nucleon center, while the shear force $s(r)$ accounts for the tangential stress that maintains mechanical stability. These quantities satisfy the conservation of forces within the nucleon which can be expressed as follows:
\begin{equation}
\label{Eq18}
   \left(\frac{2}{3} \frac{d}{dr} + \frac{2}{r} \right) s(r) + \frac{dp(r)}{dr} = 0\,, 
\end{equation}
where the first term represents the contribution from shear forces and the second term the spatial gradient of the pressure~\cite{Polyakov:2018zvc}.

The pressure and shear force distributions provide insight into the strong interaction dynamics that bind quarks and gluons within hadrons. For instance, the maximum central pressure in a nucleon exceeds $10^{35} \ \text{Pa}$, surpassing the pressure inside neutron stars~\cite{Burkert:2018bqq}. These distributions reveal fundamental information about quark confinement and the mechanical stability of nucleons, offering a new perspective on their internal structure.

The stability of the nucleon is ensured by the condition that the total internal forces vanish when integrated over the volume of the nucleon. This stability condition is expressed as:
\begin{equation}
\label{Eq19}
    \int_0^\infty dr\, r^2  p(r)= 0\,, 
\end{equation}
which ensures that the internal stresses balance each other~\cite{Burkert:2018bqq}.

The pressure and shear force distributions are directly related to the $D(t)$ GFF of the EMT. Actually, $ D(t)$ contributes to the spatial components of the EMT and is crucial for characterizing the internal mechanical properties of nucleons. It is defined from the nucleon matrix element of the EMT:
\begin{equation}
\label{Eq20}
    \langle p'| T^{ij}(0) |p \rangle = \frac{1}{m} \bar{u}(p') \left[ \frac{\Delta^i \Delta^j - \delta^{ij}\Delta^2}{4} D(t) \right] u(p), 
\end{equation}
where $\Delta = p' - p$ is the momentum transfer, $t = \Delta^2$, and $m$ is the nucleon mass~\cite{Polyakov:2002yz}.
The $D(t)$ GFF is also directly related to the internal stress tensor $T^{ij}(r)$, which can be decomposed into pressure and shear components:
\begin{equation}
\label{Eq21}
   T^{ij}(\vec{r}) = \biggl(
	\frac{r^ir^j}{r^2}-\frac13\,\delta^{ij}\biggr) s(r)
	+ \delta^{ij}\,p(r)\,,
\end{equation}
with $ r = |\vec{r}|$~\cite{Polyakov:2018zvc}. This relationship encapsulates the distribution of internal forces in terms of the GFF $D(t)$. Actually, if we define the Fourier transform of the EMT GFF $ D(t=-\bm{\Delta}^2) $ in the so-called Breit frame as 
\begin{equation}
\label{Eq22}
{\widetilde{D}(r)=}
	\int {\frac{d^3\Delta}{(2\pi)^3}}\ e^{{-i} \bm{\Delta r}}\ D(-\bm{\Delta}^2)\,,
\end{equation}
we can compute the pressure $ p(r) $ and shear forces $ s(r) $ as follows
\begin{align}
    p(r) &= p(r)=\frac{1}{6 m} \frac{1}{r^2}\frac{d}{dr} \left( r^2\frac{d}{dr}
 	{\widetilde{D}(r)} \right)\,, \label{Eq23} \\
    s(r) &= s(r)= -\frac{1}{4 m}\ r \frac{d}{dr} \left( \frac{1}{r} \frac{d}{dr}
	{\widetilde{D}(r)}\right) \,. \label{Eq24}
\end{align}

\begin{figure}[!htb]
    \centering
\includegraphics[scale=1.0]{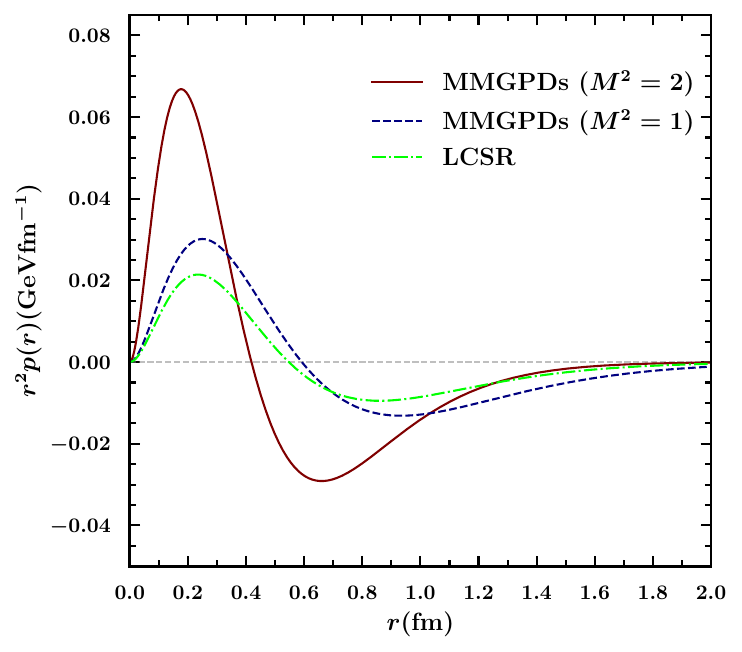}    
    \caption{Pressure distribution of the proton, $ r^2 p(r) $, as a function of the radial distance $ r $ from the center of the proton computed using the GFF $ D^Q(t) $ determined in the previous section for two values of $ M^2 $ in Eq.~(\ref{Eq17}). The LCSR result has been taken from Ref.~\cite{Dehghan:2025ncw}.}
\label{fig:Pressure}
\end{figure}

Figure~\ref{fig:Pressure} shows the results obtained for the pressure distribution $ r^2 p(r) $ that results from the interactions of the quarks in the proton versus the radial distance $ r $ from the center of the proton. In order to investigate the impact of $ M^2 $ value, we have shown the results for both $ M^2=1,2 $ GeV$ ^2 $. As can be seen, for $ M^2=2 $ GeV$ ^2 $, we find a remarkably strong repulsive pressure near the center of
the proton (up to 0.4 fm) and a binding pressure at greater
distances. In the case of $ M^2=1 $ GeV$ ^2 $, which is corresponding to the larger value of $ D^Q(0) $ in Fig.~\ref{fig:M2Scan}, we obtain smaller repulsive and binding pressures inside the proton with $ r=0.6 $ fm as the distance where the sign of pressure is changed. Note that the pressure distribution must satisfy the condition~(\ref{Eq19}). For further comparison, we also include the corresponding result from pure theoretical calculations using LCSR~\cite{Dehghan:2025ncw}. Interestingly, our result for $ M^2=1 $ GeV$ ^2 $ shows excellent agreement with the LCSR prediction.

\begin{figure}[!htb]
    \centering
\includegraphics[scale=1.0]{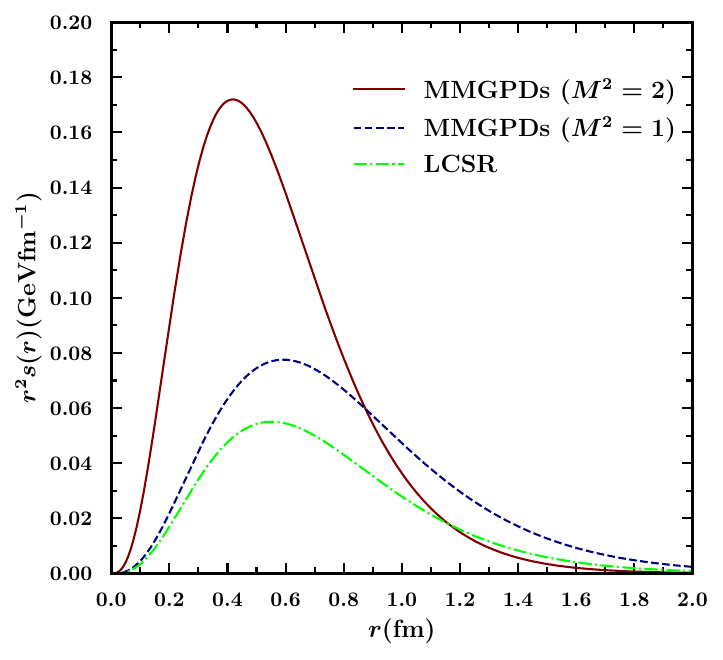}    
    \caption{The same results as Fig.~\ref{fig:Pressure} but for shear distribution in the proton.}
\label{fig:Shear}
\end{figure}

Figure~\ref{fig:Shear} shows the same results as Fig.~\ref{fig:Pressure} but for shear distribution in the proton. As one can see, choosing $ M^2=1 $ GeV$ ^2 $ leads to more moderate shear distribution than $ M^2=2 $ GeV$ ^2 $. Actually, as the value of $ M^2 $ increases, the shear distribution in the proton becomes smaller in magnitude and its peak moves to the larger values of the radial distance $ r $ from the center of the proton. This behavior is contrary to the pressure distribution in Fig~\ref{fig:Pressure} and can be attributed to the conservation condition in Eq.~(\ref{Eq18}). Note that our result for $ M^2=1 $ GeV$ ^2 $ shows again good agreement with the LCSR prediction~\cite{Dehghan:2025ncw}.

\subsection{The mechanical and mass radii}

The proton, as a composite particle, exhibits complex internal structure characterized by its charge, mass, and mechanical properties. Among these, the mechanical and mass radii provide critical insights into the spatial distribution of the proton's internal energy and pressure~\cite{Polyakov:2018zvc,Burkert:2023wzr,Cao:2024zlf,Duran:2022xag,Mamo:2021krl,Burkert:2023atx, Nair:2024fit,Yao:2024ixu}. These radii are distinct from the more commonly discussed charge radius~\cite{Goharipour:2024mbk,Karr:2020wgh,Gao:2021sml,Xiong:2023zih,Pohl:2010zza}, as they reflect different aspects of the proton's structure. The mechanical radius of the proton is derived from the spatial distribution of the EMT, which encodes information about the proton's internal pressure and shear forces. This radius is closely related to the proton's stability, as it reflects the balance between repulsive and attractive forces within the proton. The mass radius, on the other hand, describes the spatial distribution of the proton's mass, which arises primarily from the confinement of quarks and gluons. It is defined through the form factors of the EMT. 

Since both the mechanical and mass radii of the proton are related to the GFF $ D(t) $ as follows~\cite{Polyakov:2018zvc}
\begin{align}
    \langle r^2 \rangle_{\text{mech}} &= \frac{6D(0)}{\int_{-\infty}^0 D(t)\, dt} \,, \label{Eq26} \\
    \langle r^2 \rangle_{\text{mass}} &= 6 A^{\prime}(0) -\frac{3D(0)}{2m^2} \,, \label{Eq26}
\end{align}
we can now compute them (only the quark contributions) using the GFF $ D^Q(t) $
extracted in the previous section. 
In Eq.~(\ref{Eq26}), $ A(t) $ is another EMT form factor which gives knowledge
on the fractions of the momenta carried by the quark and gluon constituents of the nucleon. This form factor satisfies at zero momentum transfer the constraint $ A(0)=1 $, which means that  100\% of the hadron momentum is carried by its
constituents in the infinite momentum frame. Note also that $ A(t) $ can be calculated from the zero-skewness GPDs as follows:
\begin{equation}
\label{Eq27}
A(t)= \int_{-1}^1 dx\, x H(x,\xi=0,t)\,.
\end{equation}

\begin{figure}[!htb]
    \centering
\includegraphics[scale=1.0]{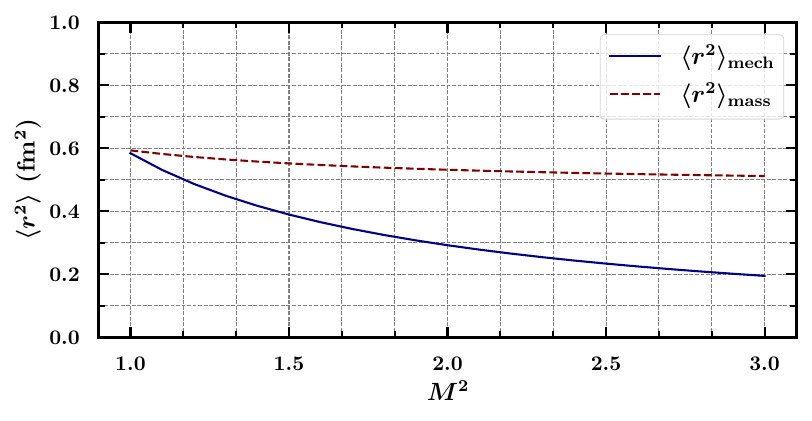}    
    \caption{The mean square mechanical and mass radii, $ \langle r^2 \rangle_{\text{mech}} $ and $ \langle r^2 \rangle_{\text{mass}} $, in terms of $ M^2 $.}
\label{fig:r2}
\end{figure}

Figure~\ref{fig:r2} shows the values obtained for the mean square mechanical and mass radii,  $ \langle r^2 \rangle_{\text{mech}} $ and $ \langle r^2 \rangle_{\text{mass}} $, in terms of $ M^2 $ values in Eq.~(\ref{Eq17}) in the interval $ 1 < M^2 < 3 $ (note that for each value of $ M^2 $ a different value of $ d_1^Q(0) $ is obtained from the fit considering the top panel in Fig.~\ref{fig:M2Scan}). According to the results obtained, choosing larger values of $ M^2 $ leads to smaller values of the mechanical radius.
For instance, we have $ \langle r^2 \rangle_{\text{mech}}= 0.58 \pm 0.04 $ and $ \langle r^2 \rangle_{\text{mech}}=0.29 \pm 0.02 $ for $ M^2=1 $ and $ M^2=2 $, respectively. The situation is the same for the mass radius with  the difference that it is less sensitive to change in $ M^2 $. In this case, we have $ \langle r^2 \rangle_{\text{mass}}= 0.59 \pm 0.02 $ and $ \langle r^2 \rangle_{\text{mass}}=0.53 \pm 0.02 $ for $ M^2=1 $ and $ M^2=2 $, respectively. An interesting point is that similar values have been obtained for the mechanical and mass radii of the proton  at $ M^2=1 $ GeV$ ^2 $. However, considering the fact that the mechanical radius is particularly sensitive to the gluonic contributions to the proton's structure, as gluons dominate the energy-momentum tensor in QCD, it seems that a larger value like $ M^2=2 $ would be a more realistic choice (note that in the present analysis we just have determined the quark contributions in the GFF $ D(t) $ and, hence, in other related quantities). 

\begin{figure}[!htb]
    \centering
\includegraphics[scale=0.5]{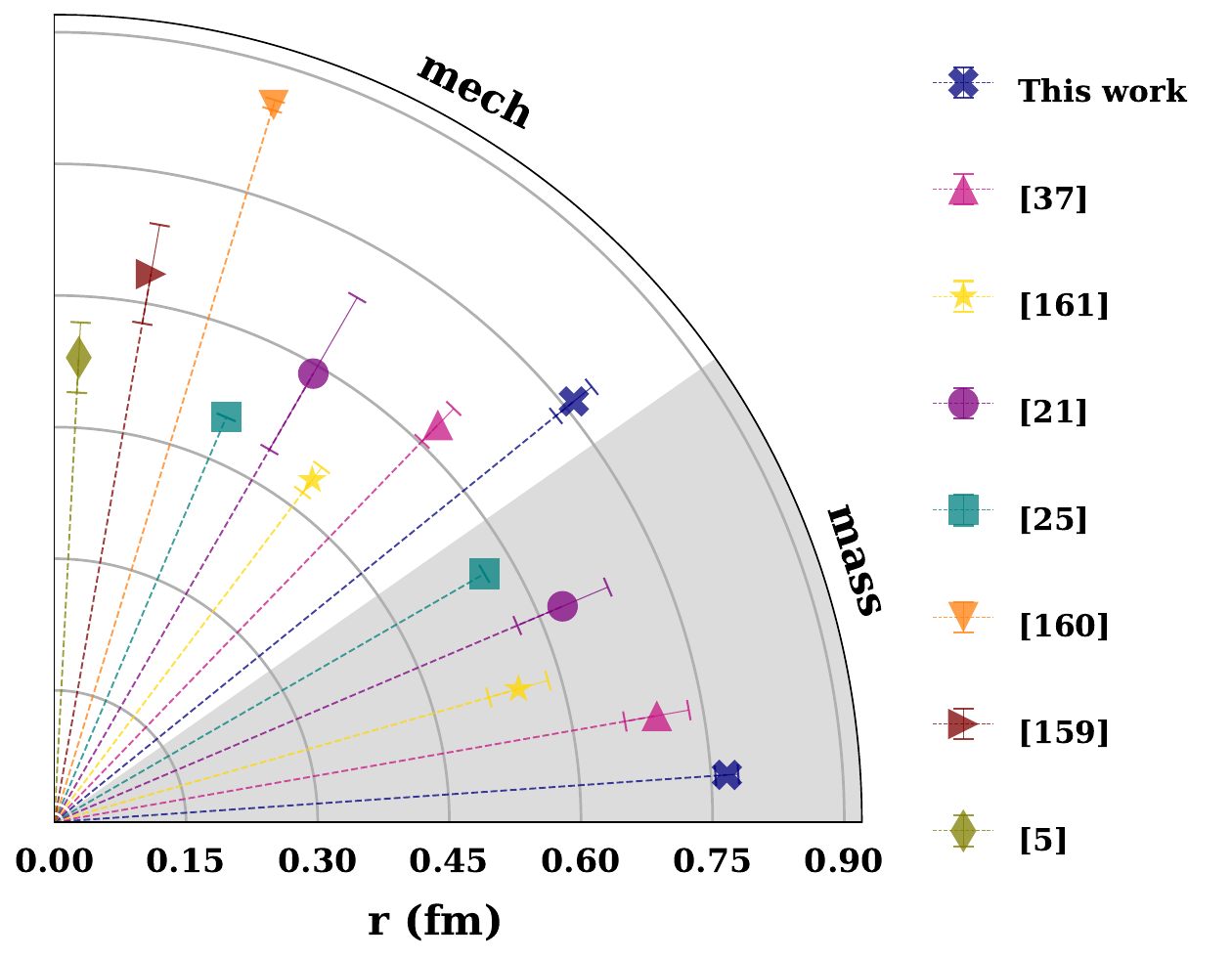}    
    \caption{A comparison  between our results obtained for the mechanical and mass radii of the proton (\( M^2 = 1 \, \text{GeV}^2 \)) and the corresponding ones obtained from other studies. See text for more details.}
\label{fig:r2Comp}
\end{figure}

Figure~\ref{fig:r2Comp} compares our results for the proton's mechanical and mass radii at \( M^2 = 1 \, \text{GeV}^2 \) with those obtained from other studies. These include results from LCSR~\cite{Dehghan:2025ncw,Azizi:2019ytx}, lattice QCD~\cite{Hackett:2023rif}, light-front quark-diquark model~\cite{Choudhary:2022den}, the light-front quantization approach by the BLFQ Collaboration~\cite{Nair:2024fit}, and a symmetry-preserving analysis by Yao \textit{et al.}~\cite{Yao:2024ixu}. It is important to note that all results correspond to the quark contributions to the mechanical and mass radii.

As shown in the figure, the results span a wide range: \( 0.488-0.854 \, \text{fm} \) for \( r_{\text{mech}} \) and \( 0.55-0.768 \, \text{fm} \) for \( r_{\text{mass}} \). A simple combination (average) of all results yields \( r_{\text{mech}} = 0.623 \pm 0.018 \, \text{fm} \) and \( r_{\text{mass}} = 0.642 \pm 0.019 \, \text{fm} \). For a comprehensive review of recent results on the proton's mechanical and mass radii from various phenomenological and theoretical studies, we refer readers to Ref.~\cite{Goharipour:2025yxm}.
%

%

\section{Summary and conclusion}\label{sec:five}

In this study, we have investigated the GFFs of the proton, with a particular focus on the D-term, which encodes critical information about the internal stress distribution, pressure, and shear forces within the nucleon. By constructing skewness-dependent GPDs using the double-distribution representation and leveraging zero-skewness GPDs derived from elastic scattering data, we extracted the quark contribution to the $ D(t) $ GFF of the EMT through an analysis of the CFFs data. Our results provide new insights into the mechanical properties of the proton, including its pressure and shear force distributions, as well as its mechanical and mass radii.

A key finding of this work is the remarkable value obtained for $ D^Q(0) $, which contrasts with results derived from analyses of DVCS observables. This discrepancy highlights the importance of considering a wider range of momentum transfer $ t $, as elastic scattering processes cover larger $ t $ values compared to DVCS, which is limited to smaller $ t $. Our results suggest that future hard-exclusive data from facilities such as Jefferson Laboratory, COMPASS at CERN, and the Electron-Ion Collider (EIC) may yield more significant values for the EMT form factor $ D^Q(t) $ than currently available data. However, further validation of our approach using original DVCS data is necessary to include more data points and constrain the fit parameters more rigorously. This will be the focus of a follow-up study.

The pressure and shear force distributions within the proton were also explored. For $ M^2=2 $ GeV$ ^2 $ (see the text for more details), we observed a strong repulsive pressure near the proton's center (up to 0.4 fm) and a binding pressure at larger distances. In contrast, for $ M^2=1 $ GeV$ ^2 $, which corresponds to a larger $ D^Q(0) $, the repulsive and binding pressures were smaller, with the pressure sign change occurring at $ r=0.6 $ fm. The shear force distribution exhibited an inverse trend, with larger $ M^2 $ values leading to smaller magnitudes and a shift in the peak toward larger radial distances. These behaviors are consistent with the conservation conditions governing the pressure and shear distributions.

Additionally, we calculated the mean square mechanical and mass radii $ \langle r^2 \rangle_{\text{mech}} $ and $ \langle r^2 \rangle_{\text{mass}} $ as a function of $ M^2 $. Our results indicate that larger $ M^2 $ values correspond to smaller mass and mechanical radii, especially for the last one. For instance, we obtained $ \langle r^2 \rangle_{\text{mech}}= 0.58 \pm 0.04 $ and $ \langle r^2 \rangle_{\text{mass}}= 0.59 \pm 0.02 $ for $ M^2=1 $ GeV$ ^2 $ as well as $ \langle r^2 \rangle_{\text{mech}}=0.29 \pm 0.02 $ and $ \langle r^2 \rangle_{\text{mass}}=0.53 \pm 0.02 $ for $ M^2=2 $ GeV$ ^2 $. These findings underscore the sensitivity of the mechanical radius to the choice of $ M^2 $ and provide valuable constraints for theoretical models of nucleon structure.

In conclusion, this work advances our understanding of the proton's internal mechanical properties by leveraging skewness-dependent GPDs and CFF data. The extracted D-term and associated mechanical observables offer new perspectives on the distribution of pressure and shear forces within the proton, as well as its mechanical and mass radii. Our findings highlight the importance of continued investigations into the GFFs and their role in elucidating the fundamental structure of matter.

%
\section*{ACKNOWLEDGEMENTS}
The authors thank V. D. Burkert, Krešimir Kumerički, and Yuxun Guo for useful comments and discussions. M.~Goharipour is thankful to the School of Particles and Accelerators and School of Physics, Institute for Research in Fundamental Sciences (IPM), for financial support provided for this research.  F. Irani and K. Azizi are thankful to Iran National Science Foundation (INSF) for financial support provided for this research under grant No. 4033039.

%

%

\end{document}